\documentstyle[12pt,epsf]{article}
\newcommand{\beq}{\begin{equation}}
\newcommand{\eeq}{\end{equation}}
\input epsf
\begin{document}
{\title{QCD-Motivated Pomeron and High Energy
Hadronic Diffractive Cross Sections.}
\vskip 1cm
\author{ V.V.Anisovich$^{1)}$, L.G.Dakhno$^{2)}$ and V.A.Nikonov$^{3)}$}
\date{ }
\maketitle

\vskip 1cm
\begin{center}
St.Petersburg Nuclear Physics Institute,\\
Gatchina, St.Petersburg 188350, Russia\\
$^{1)}$ anisovic@lnpi.spb.su\\
$^{2)}$ dakhno@hep486.pnpi.spb.ru\\
$^{3)}$ nikon@rec03.pnpi.spb.ru
\end{center}
\bigskip

\begin{abstract}
Soft diffractive cross sections in  $pp$ (or $\bar pp$), $\pi p$
and $\gamma p$ collisions are calculated using QCD-motivated
pomeron. The $s$-channel unitarization of the amplitude is performed
by the eikonal approach, and in the framework of the quark structure
of hadrons the colour screening is taken into account.
The performed description of the diffractive processes has led to the
parameters of the bare pomeron $P$ which occurred to be close to
the corresponding parameters of the Lipatov's pomeron [1].
Parameters of the bare pomeron and those of the three-reggeon block
 $PGG$ ($G$ is a reggeized gluon) has been fixed by the data at moderately
 high energies;
 for superhigh energies the predictions are made. The intercept of the
bare pomeron is found to be in remarkable agreement with the low-$x$
data for deep inelastic scattering.
An evaluation of the effective colour transparency radius is done
 which can be used in hadron-nuclei reactions.
\end{abstract}

\bigskip

In the latest decade the phenomenon of a pomeron  attracts
much attention; still, the pomeron structure remains enigmatic
in many respects.
 A perpetual item of the  agenda is the description of the
pomeron in  terms consistent with QCD (see, for example, ref. [2] and
references therein). Although the growth of the QCD
coupling constant at large distances prevents
the direct use of  perturbative QCD for the description
of soft processes,         it looks as if  the pomeron, being a
small-size object, provides us with an exception. In this case the bare
pomeron of  perturbative QCD [1,3] seems to be an appropriate object
for use as a guide to  the description of diffractive processes
at high and superhigh energies.

The behaviour of amplitudes for hadronic diffractive processes is determined
by singularities of the $t$-channel partial waves in the complex plane
of the angular momentum $j$, and the singularity in the channel with vacuum
quantum numbers (pomeron) dominates at high energies.
The pomeron of the SU(N) Yang-Mills theory in the leading logarithmic
approximation appears as a composite system of reggeizied gluons while
the corresponding partial waves have a fixed root singularity at
$j=1+\Delta_{BFKL}$, where $\Delta_{BFKL}=(g/\pi)^{2}N\ln2$ (this
is so called BFKL-pomeron[3], $\Delta\cong 0.5$ at $N=3$).
Application of QCD-pomeron in the phenomenological calculations makes it
urgent to consider in the gluon ladder the virtual momenta which are close
to  those of the leading logarithmic approximation.
In ref. [1] the virtualities of such a type have been taken into account
using certain boundary condition as well as the constraint ensured by the
asymptotic freedom of the SU(3) theory (QCD).
 The pomeron obtained in this way [1] is an infinite
set of poles in the region $1<j\le1+\Delta$, and there exists a
 constraint on the intercept of the leading pole: $\Delta \ge 0.3$.

The dynamics of the pomeron reveals itself  in the
small-$x$ deep inelastic proceses: at moderate $Q^2$ the structure function
$F_2$ should behave as $x^{-\Delta}$. The global fit which is performed in
ref. [4] and is based on the world data as well as on the new measurements
of $F_2(ep)$ by ZEUS
and H1-collaborations [5] provides the value $\Delta=0.3$. The numerical
solution of the BFKL-equation [6]
 mimics $x^{-0.3}$ behaviour, although it gives a bit larger value for
 $\Delta$ which is close to 0.5.

There are reasons to believe that a
successful description of soft diffractive
processes is possible with a pomeron whose  characteristics are close
to those of perturbative pomeron. The point is that the phenomenological
pomeron used at moderately high energies for the description of the
diffractive processes is an almost point-like system (see ref. [7] and
references
therein). This is supported by a small value of $\alpha '_P$ as well as by
large masses of the resonances which are candidates for glueballs,
$M_G \sim 1.5 \div 2.2$ GeV. As a result
 the integration  in the gluon ladder of the pomeron
is carried out over large momenta, where the running coupling constant
$\alpha_s$ is not large. The minijet picture [8], which is successful
in the description of the multiparticle production in hadron-hadron
collisions, also requires comparatively large momenta in the gluon ladder.

The attempt to work with the BFKL-like pomeron but with small
 $\Delta$ for the description of $\pi p$ and $pp$-scatterings at
moderately high energies has been made in ref. [9]. The pomeron,
being a gluon ladder  (Fig. 1$a$), has two types
of couplings with quarks of the scattered hadron:
with one quark (Fig. 1$b$) and with
two of them (the last type of coupling occurs actually via the three-reggeon
vertex $PGG$, where $P$ and $G$ are pomeron and reggeized gluon,
correspondingly, see Fig. 1$c$). These two couplings cancel
each other at small interquark
distances causing colour screening. The small size of the pomeron reveals
itself in a small magnitude for the colour screening radius, $r_{cs}$, --
this fact allows to understand why  the additive quark model does
work at moderately high energies.
Moreover, colour screening effects are able to explain the deviation from
additivity in $\sigma_{tot}(\pi p)$ and $\sigma_{tot}(pp)$ at $\sqrt {s}
\sim 20$ GeV [9] ($\sigma_{tot}(pp)/\sigma_{tot}(\pi p)\simeq 1.6$). However,
the growth of the total $pp/\bar pp$ cross sections at $\sqrt {s} > 30$ GeV
requires the value $\Delta >0$ [10]. Still, the value $\Delta=0.08$, though
good in fitting the data, creates  problems with $s$-channel unitarity.
To satisfy the $s$-channel unitarity one should take into account  the
rescatterings in the $s$-channel.
This procedure does not converge
at small $\Delta$ -- of the order of 0.1: the more rescatterings
 are taken into
account the larger the value of $\Delta$  needed for the input pomeron.
The description of  $pp$,  $\pi p$ and $\gamma p$ diffractive cross sections
$\sigma_{tot},\sigma_{el},\sigma_{DD}^{single}$ and $\sigma_{DD}^{double}$
presented in this paper is performed with the full set of $s$-channel
rescatterings taken into account in the  eikonal
approximation. As a result, we got the value $\Delta=0.29$: this magnitude
is on the lower boarder of the constraint for the Lipatov's  pomeron
and coincides with the value obtained in the deep
inelastic HERA experiments [5].

Further presentation is organized as follows: first of all, we give the
 formulae for the diffractive cross sections, then we make some comments
 on their derivation.
 The following formulae
describe total, elastic and diffractive dissociation cross sections
of colliding hadrons $A$ and $B$:
\beq
\sigma_{tot}(AB)=2\int d^2b \int dr'\varphi^2_A(r')dr''\varphi^2_B(r'')
\left [1-\exp{(-\frac{1}{2}\chi_{AB}(r',r'',b)})\right],
\eeq
\beq
\sigma_{el}(AB)=\int d^2b
\left \{\int dr'\varphi^2_A(r')dr''\varphi^2_B(r'')
[1-\exp{(-\frac{1}{2}\chi_{AB}(r',r'',b)})]\right\}^2
\eeq
\beq
\sigma^{single}_{DD(B)}(AB)
+\sigma_{el}(AB)=
\int d^2 b \int dr'\varphi^2_A(r')dr''\varphi^2_B(r'')
d\tilde{r}'\varphi^2_A(\tilde{r}')
\eeq
$$\left [1-\exp{(-\frac{1}{2}\chi_{AB}(r',r'',b)})\right]
\left [1-\exp{(-\frac{1}{2}\chi_{AB}(\tilde{r}',r'',b)})\right],$$
\beq
\sigma_{HD}(AB)=
\sigma_{el}(AB)+\sigma^{single}_{DD(A)}(AB)+\sigma^{single}_{DD(B)}(AB)
+\sigma^{double}_{DD}(AB)=
\eeq
$$\int d^2b
\int dr'\varphi^2_A(r')dr''\varphi^2_B(r'')
\left [1-\exp(-\frac{1}{2}\chi_{AB}(r',r'',b))\right]^2.$$
The last expression, $\sigma_{HD}(AB)$, stands for full hadron diffraction.
Here $dr\varphi^2_{A,B}(r)$ are the quark densities of  colliding hadrons
$A$ and $B$ which depend on the transverse coordinates:
\beq
dr \varphi^2_{\pi}(r)=d^2r_1d^2r_2\delta ^2(\vec {r}_1+\vec {r}_2)
\varphi^2_{\pi}(r_1,r_2),
\eeq
$$
dr \varphi^2_p(r)=d^2r_1d^2r_2d^2r_3\delta ^2(\vec {r}_1+\vec {r}_2
+\vec {r}_3)
\varphi^2_p(r_1,r_2,r_3);$$
$r_i$ is the transverse coordinate of a quark, and the wave function squared
$\varphi_A^2$ has been  integrated over
longitudinal variables. Proton and pion quark densities are determined
using correspopnding form factors; they are presented in
Appendix.
The profile-function $\chi_{AB}$ corresponds to the
interaction of quarks via pomeron exchange as follows:
\beq
\chi_{AB}(r',r'',b)=\int db'db''\delta^2(b-b'+b'')S_A(b',r')S_B(b'',r'').
\eeq
Functions $S_{A,B}$ stand for the pomeron-quark interaction;
they are
determined by the diagrams with different couplings of the pomeron with
quarks as is written below:
\beq
S_{\pi}(\vec {r},\vec {b})=\rho(\vec {b}-\vec {r}_1)+\rho(\vec {b}-\vec
{r}_2)-
2\rho(\vec {b}-\frac{\vec {r}_1+\vec {r}_2}{2})\exp(-\frac{(\vec {r}_1-
\vec {r}_2)^2}{4r^2_{cs}}),
\eeq
$$S_p(\vec {r},\vec {b})=
\Sigma_{i=1,2,3}\rho(\vec {b}-\vec {r}_i)-
\Sigma_{i\ne k}
\rho(\vec {b}-\frac{\vec {r}_i+\vec {r}_k}{2})\exp(-\frac{(\vec {r}_i-
\vec {r}_k)^2}{4r^2_{cs}}).$$
The term $\rho(\vec {b}-\vec {r}_i)$ describes the diagram where the pomeron
couples to one of the hadron quarks (Fig. 1$b$) while the terms
proportional to $\exp(-r^2_{ij}/r^2_{cs})$ are related to the diagram 1$c$
with the pomeron which couples to two quarks of the hadron.
This diagram is a
three-reggeon graph where $G$ is the
reggeized gluon. Functions $S_{\pi}$ and
$S_p$
tend to zero as $|\vec {r}_{ij}| \to 0$: this is the colour
screening phenomenon  inherent to the Lipatov's pomeron. For the
sake of convenience,
we perform calculations in the cms of the colliding quarks, supposing that
hadron momentum is shared equally between its quarks. Then
\beq
\rho(b)= \frac {g}{4\pi(G+\frac{1}{2}\alpha'_P\ln{s_{qq}})}
 \exp\left [-\frac{b^2}{4(G+\frac{1}{2}\alpha'_P\ln{s_{qq}})}\right],
\eeq
where the vertex $g$ depends on the energy squared
of the colliding quarks, $s_{qq}$:
\beq
g^2=g_0^2+g_1^2\left ( \frac{s_{qq}}{s_0} \right )^{\Delta}     .
\eeq
Such a parametrization of $g^2$ corresponds to the two-pole presentation
of the QCD-motivated pomeron with $j_0=1$ and $j_1=1+\Delta$. Here and below
$s_0=1$ GeV$^2$.

Now let us make comments on the formulae (1)-(4).
Eqs. (1)-(4) in the case of a pion beam are obtained summing the
diagrams like Fig. 1$d$,  where all the possible meson states
$M_i$ and $M_j$ are taken into account.
The assumption that a full set of meson states $M_i$ corresponds to
a full set of the quark-antiquark states leads to the diagram 1$e$, and just
this type
of diagrams with the quark intermediate states is reproduced by eqs. (1)-(4).
Analogously the diagrams like Fig.1 $f$  are taken into account in eqs.
(1)-(4) in the case of a proton beam.

The diffractive cross sections (1)-(4) for a meson beam were obtained in
ref. [9], while for the proton beam the final formulae were reported in
ref. [11]; their derivation will be published elsewhere.

Eqs. (1)-(4) depend on the transverse coordinates of quarks,
 though the
original expressions depend on the fractions of the momenta of the colliding
hadrons carried by the quark, $x_i$. In the functions $S$ we put $x_i=1/2$
for a meson and $x_i=1/3$ for the proton, in other words we assume that
hadron wave functions $\varphi_{\pi}(\vec {r},x)$ and
$\varphi_p(\vec {r},x)$ select the mean values of $x_i$ in the interaction
blocks. A wide range of wave functions obey this asumption, for example,
the wave functions of  quark spectroscopy. But the
situation with  the diagram of Fig. 1$c$ is more
complicated. One should perform an integration over the part of the
energy carried by reggeized gluons and pomeron: this spreads
the $x_i$'s of the interacting quarks. However, if the intercept
of the reggeized gluon $\alpha_G(0)$ is near 1 (it is actually the
requirement of the BFKL pomeron), then $x_i$'s can be considered as
frozen. We have checked by numerical calculations that this assumption
works at $0.8<\alpha_G(0)<1$ for realistic pion and proton wave
functions. In due course, in eq. (9) we put $s_{qq}=s/6$
for $\pi p$ collision and $s_{qq}=s/9$ for $pp$.

Eqs. (1)-(4) can be used at small momentum transfers where
real parts of the amplitudes are small. Hence we neglected the
signature factor
of the bare pomeron, but it can be easily restored. We shall
return to this point later on.

Eqs. (1)-(4) are written in the eikonal approximation for composite systems.
It is well known that certain correlations are missed in this procedure.
Let us elucidate what type of correlations is taken into account and
what is neglected in the developed approach. Interactions of quarks
of the same hadron and the transitions of these quarks into the excited
hadron states are included in the diagrams of type 1$d$, and the
completeness condition for the quark states results actually in multiple
interactions of the "frozen" quark state (diagram
of Fig. 1$e$). But the $t$-channel
gluon interactions are taken into account only in their simplest form - as a
set of non-interacting pomeron exchanges. Including  pomeron-pomeron
interactions is actually the problem of finding a solution
which satisfies simultaneously both $t$- and $s$-channel unitarity
conditions. Attempts to solve it have  been intensified recently [12].
Calculations performed here provide arguments that eikonal diagrams with
the found pomeron parameters play a decisive role in the formation of
diffractive processes at present high and superhigh energies.
 This point will be discussed below in a more detail.

Let us discuss  results of the calculation.
Total and elastic $pp$ (or $p\bar p$) and $\pi p$ cross sections
are shown
in Fig. 2$a,b$. The parameters of the input pomeron have been obtained
in the fitting procedure in a broad energy range ($\sqrt {s}=23.7\div 1800$
GeV); they are as follows:
\beq
g_0^2=7.914\;{\rm mb},\;\;\;g_1^2=0.179\;{\rm mb},\;\;\;r_{cs}=0.18\;
{\rm fm};
\eeq
$$\Delta=0.29,\;\;\;G= 0.167\; {\rm (GeV/c)}^{-2},\;\;\alpha'_P=0.112\;
{\rm (GeV/c)}^{-2}.$$
The wave functions $\varphi_{\pi}$ and $\varphi_p$ are chosen to satisfy
pion and proton form factors at $|q^2| \le 1$ GeV$^2$ in the framework
of the quark model.

At asymptotic energies the total cross sections $\sigma_{tot}(pp)$ and
$\sigma_{tot}(\pi p)$ calculated with the parameters (10) are shown in
Fig. 2$a$; at superhigh energies they
increase as $0.32\ln^2s$ mb.
The growth with energy of the elastic cross sections, $\sigma_{el}(pp)$ and
$\sigma_{el}(\pi p)$, calculated with eq. (2) is presented in Fig. 2$b$.
At superhigh energies they grow as $0.16\ln^2s$ mb.
Total cross sections calculated with the parameters (10) can be fitted
at $\sqrt{s} \ge 100$ GeV, within 5\% accuracy, by the following
expressions:
\beq
\sigma_{tot}(pp)=1.75 +2.27\ln (s/s_0)+0.32\ln ^2 (s/s_0),
\eeq
$$\sigma_{tot}(\pi p)=4.93-6.19\ln (s/s_0)+0.32\ln ^2 (s/s_0).$$
and elastic cross sections  can be fitted in the following form:
\beq
\sigma_{el}(pp)=-6.13+0.797\ln (s/s_0)+0.16\ln ^2 (s/s_0),
\eeq
$$\sigma_{el}(\pi p)=3.05-1.38\ln (s/s_0)+0.16\ln ^2 (s/s_0).$$
In eqs. (11) and (12) numerical coefficients are given in mb, $s_0=1$
GeV$^2$.

For LHC energy ($\sqrt{s}=16$ TeV) our predictions, which relate to the bare
pomeron parameters given by eq. (10), are as follows:
              $\sigma_{tot}(pp)=131$ mb and $\sigma_{el}(pp)=41$ mb.

The slopes $B$ for the elastic $pp(\bar pp)$ and $\pi p$ cross
sections $d\sigma_{el}/dq^2 \sim \exp (-Bq^2)$ are shown in Fig. 2$c$: the
calculated values describe well the data.

The ratio of the total cross sections
$\sigma_{tot}(pp)/\sigma_{tot}(\pi p)$ tends to unity at
far asymptotic energies.
In this limit $\sigma_{el}(pp)/\sigma_{tot}(pp) \to 1/2$ and
$\sigma_{el}(\pi p)/\sigma_{tot}(\pi p) \to 1/2$ ( Pumplin's limit
[14]). These limit
magnitudes originate because of the disappearance
of the colour screening radius at superhigh energies: effective colour
screening radius, which is different for $pp$ and $\pi p$ collisions,
tends to zero at asymptotic energies.

As was mentioned above, the restoration of the pomeron signature factor
(or crossing symmetry of the amplitude) can be easily done, and this allows
one to calculate $\rho=Re\,A/Im\,A$. The result is shown in Fig. 2$d$: the
description of the data for $pp(\bar pp)$ collisions is rather reasonable.

Now let us discuss the process of the diffraction dissociation. First,
there is a problem of the definition of $\sigma^{single}_{DD}$, for two
mechanisms contribute to the measured diffractive dissociation cross
section: one is the dissociation of the colliding hadron, see Fig. 3$a$,
and another involves partly dissociated pomeron, Fig. 3$b$. Eqs. (3) and (4)
describe the hadron dissociation only, the calculated cross section for the
proton dissociation is presented in Fig. 4$a$. The difference of the
measured value of $\sigma^{single}_{DD}(\bar pp)$ and the calculated
one  provides
just the cross section for the partly dissociated pomeron which is
determined by the three-pomeron diagram: this difference is shown in
Fig. 4$b$. It should be pointed out that the developped approach allows to
calculate the other characteristics of the diffraction dissociation of
the colliding particles, namely, the $M^2$- and $t$-dependences. However
such a study  is beyond our present consideration.

In Fig. 5 the total cross section $\sigma_{tot}(\gamma p)$ is presented,
together with the available experimental data
obtained by ZEUS and H1 (see ref. [15] and references therein). The
calculations have been
performed in the framework of the hypothesis that a hadronization of the
photon is governed by the vector dominance model, $\gamma \to V$, and
the wave functions of the vector mesons
 ($\rho$ and $\omega$) are
 almost the same as for the pion,
$\psi_V \simeq \psi_{\pi}$, under a suggestion to obey the
$SU(6)$-symmetry. However,
 it is worth noting that total cross section is poorly
sensitive to the details of the wave function. More sensitive are the
two-particle reactions $\gamma p \to \rho p$ and $\gamma p \to \omega p$.
In Fig. 5 we demonstrate the  cross sections for these
reactions calculated under the assumptions that $V=\rho+\omega$
and $\sigma(\gamma p \to \rho p)=\sigma(\gamma p \to \omega p)$.

One may be rather sceptical about the eikonalization procedure.
Here we would like to present arguments why
the eikonal is a good approach in a transitional region,
when diffractive cross sections start to increase, reaching their asymptotic
regime, $\sigma \sim \ln^2 s$.
Diffractive $\log s$-physics is related to the interaction which leads to
a black disk in the impact parameter space
with the radius increasing as $\ln s$. The important point is that this
radius is mainly
determined by the interaction component which increases the most rapidly
with $s$.
Such a component is just one-pomeron exchange.
The arguments are based on the fact that the growth of the pomeron
interaction obtained in the description of the experimental
 data is strongly delayed: the parameters of the bare pomeron
provide $g^2 \sim 1+(s/3 \cdot 10^5 \cdot GeV^2)^{0.29}$.

Consider as an example the diagram
where two pomerons interact with each other; then
 the four-pomeron block emerges but with much less energy
per one pomeron,
$s_i \sim ms^{1/2}$. Because of that the polynomial growth with energy
of the four-pomeron block
comes much later - when the black disk in the impact parameter space has
been already created by one-pomeron exchanges. The black disk cannot become
more black due to additional interactions: geometrical parameters of the
disk and its rim have been already fixed by one-pomeron interactions, which
are involved in the eikonalization procedure.

Such are the arguments why diffractive processes, which are determined by the
disk geometry, can be described by the eikonal.
Still, it would not be correct  to declare
 that the four-pomeron block is not significant
at superhigh energies: it is  important when the diagram cutting is done,
i.e. in the formation of the inclusive spectra in the central
region or in multiparticle processes but not in the soft diffractive
processes.

There are two important points in the log-physics when discussing
the approach of the cross section to its asymptotic regime.
The first one concerns the number
of independent black disks which are created by the interaction before the
asymptotics.
In the quark model the number of disks is determined by the
number of constituent quarks. At intermediate energies, when black disks
are separated from each other, the screening of disks by each other
is very important,
and their attenuation leads to the delay of the coming $\ln ^2 s$-asymptotics.
Attenuation is strongly dependent on the number of
interaction sources in the hadron. The second point concerns the type
of attenuation -- now we speak about colour screening:
for example, in the squeezed quark-antiquark system the colour screening
cancels the interaction totally, while the geometrical screening cancels it
by one half only. This is the reason why we payed much attention to the
colour screening using Lipatov's pomeron as a guide.

The idea of colour screening which is realized here on the basis of the
gluon structure of the pomeron allows one to introduce the effective
colour screening radius which can be applied for the calculation of
a hadron-nucleus reaction in a way as it has been done in
ref. [16]. For this purpose we introduce the following colour screening
profile factor (for the case of a pion beam):
\beq
\chi_{\pi}(r)=N\,\int d^2b dr''\varphi^2_p(r'')
\left [1-\exp{(-\frac{1}{2}\chi_{\pi p}(r,r'',b)})\right].
\eeq
Here $N$ is a normalization factor which is chosen to satisfy the
requirement $\chi_{\pi}(r \to \infty)=1$.
Physical meaning of this function is simple: $\chi_{\pi}(r)$
describes the pion-proton interaction depending on the pion
interquark distance. The concept of colour screening means
a disappearance of interaction when the quarks are close to each other,
namely, at
the distance which is less than the colour screening radius.
After the integration over  impact parameter and      proton coordinates,
 with the parameter values given in eq. (10), we got $\chi_{\pi}(r)$
 which is presented in Fig. 6: dashed curve stands for the energy
 $\sqrt{s}=546$ GeV
(it should be pointed out that the calculations performed in the
interval $30<\sqrt{s}<1800$ GeV differ
very little from each other); dotted curve is calculated at far
asymptotic energies $s=10^{30}$ GeV$^2$; solid curve stands for
 $\chi_{\pi}(r)$ parametrized in the form
$1-\exp(-r^n/r_{cs,eff}^n)$ with $n=2.86$ and $r_{cs,eff}=0.281$ fm.

The last item which is to be discussed here is how the
calculations performed here relate to the description of the diffractive
processes at moderately high energies in the framework of the one-pomeron
exchange. This has been done in ref. [9]  for the energy range
$\sqrt{s}=15 \div 25$ GeV where the $pp$-scattering amplitude
has been described
with the exchange of the effective pomeron; the corresponding
impulse approximation diagram is shown in Fig. 7$a$. In the approach
used in this paper the set of diagrams shown in Fig. 7$b$
is a counterpart of Fig. 7$a$. In Fig. 8 the cross section related to
the diagrams shown in Fig. 7$b$ is presented depending on the
energy; let us denote it as $Im\,A_{qq}$. This magnitude which
can be called an
effective quark-quark cross section behaves at superhigh energies as
$Im\,A_{qq} \simeq \ln^2 s$. However at low energies the value $Im\,A_{qq}$
is close to the parameters found in ref. [9], namely, $Im\,A_{qq}=6$ mb at
$\sqrt{s}=23.7$ GeV, while the magnitude found in ref. [9] is 5.5 mb. Thus,
the calculation performed here is sewed rather well with the picture of
the effective one-pomeron exchange used at moderately high energies.

Summing up the results of calculations we should underline that the
parameters of the bare pomeron are close to those of Lipatov's pomeron [1].
They are also close to the pomeron parameters found in the low-$x$
deep inelastic experiments [4,5].

All calculations have been performed using the program of Monte-Carlo
simulation VEGAS [17].

The authors are indebted to N.A.Kivel, L.N.Lipatov and M.G.Ryskin
for useful discussions. One of us (VVA) is grateful
to R.S.Fletcher, T.K.Gaisser and T.Stanev for the initiating discussions
and hospitality during the visit in Delaware University.
This research has been supported by International Science Foundation,
Grants R-10000 and R-10300.

\newpage
\begin{appendix}
\section{Proton and pion quark densities}
\vskip 1cm

We describe proton and pion form factors as a sum of three exponentials:
$$F_{\pi}=\frac{1}{8\pi^{3/2}}\left[\frac{a^2_{\pi}}{(2\gamma_{\pi})^{3/2}}
e^{-\frac {\gamma _{\pi}}{8}q^2}+
\frac{2a^2_{\pi}b^2_{\pi}}{(\gamma_{\pi}+\delta _{\pi})^{3/2}}
e^{-\frac {\gamma _{\pi}\delta_{\pi}}{4(\gamma _{\pi}+\delta _{\pi})}q^2}
+\frac{b^2_{\pi}}{(2\delta_{\pi})^{3/2}}
e^{-\frac {\delta _{\pi}}{8}q^2}\right],$$
$$F_p=\frac{1}{24\sqrt{3}(2\pi)^3}\left[\frac{a^2_p}{8\gamma_p^3}
e^{-\frac {\gamma _p}{3}q^2}+
\frac{2a^2_pb^2_p}{(\gamma_p+\delta _p)^3}
e^{-\frac {2}{3} \frac {\gamma _p\delta_p}{4(\gamma _p+\delta _p)}q^2}
+\frac{b^2_p}{8\delta_p^3}
e^{-\frac {\delta _p}{3}q^2}\right],$$
with    $a_{\pi}=61.052,\;\;b_{\pi}=14.290,\;\;\gamma_{\pi}=22.222,
\;\;\;\delta_{\pi}=2.533$ and   $a_p=7307.8,\;\;b_p=502.4,\;\;
\gamma_p=11.36,\;\;\;\delta_p=2.24$. These magnitudes correspond to the
mean value of the radius squared of pion and proton:
 $<r^2_{\pi}>=10$  (GeV/c)$^{-2}$ and
$<r^2_p>=17$ (GeV/c)$^{-2}$.

In the  $r$-representation pion and proton form factors are as follows:
$$F_{\pi}(q^2)=\int d^2r_1 d^2r_2 \delta(r_1+r_2)
\varphi^2_{\pi}(r_1,r_2)e^{iqr_1},$$
$$F_p(q^2)=\int d^2r_1 d^2r_2 d^2r_3\delta(r_1+r_2+r_3)
\varphi^2_p(r_1,r_2,r_3)e^{iqr_1}.$$
Here $\varphi^2_{\pi}=\Sigma^3_{i=1}\;A_i \exp[-a_i(r_1^2+r_2^2)]$ and
 $\varphi^2_p=\Sigma^3_{i=1}\;D_i \exp[-d_i(r_1^2+r_2^2+r_3^2)]$
with the parameters:
$$A_1=0.0080902,\;\;A_2=0.044519,\;\;A_3=0.10104,\;\;$$
$$a_1=0.045000,\;\;\;a_2=0.21990,\;\;\;a_3=0.39479,$$
$$D_1=0.00026655,\;\;D_2=0.0015684,\;\;D_3=0.0041892,\;\;$$
$$d_1=0.044401,\;\;d_2=0.13471,\;\;d_3=0.22502.$$
\end{appendix}

\newpage
\begin{figure}
\begin{center}
\mbox{\epsfxsize=13cm \epsfbox{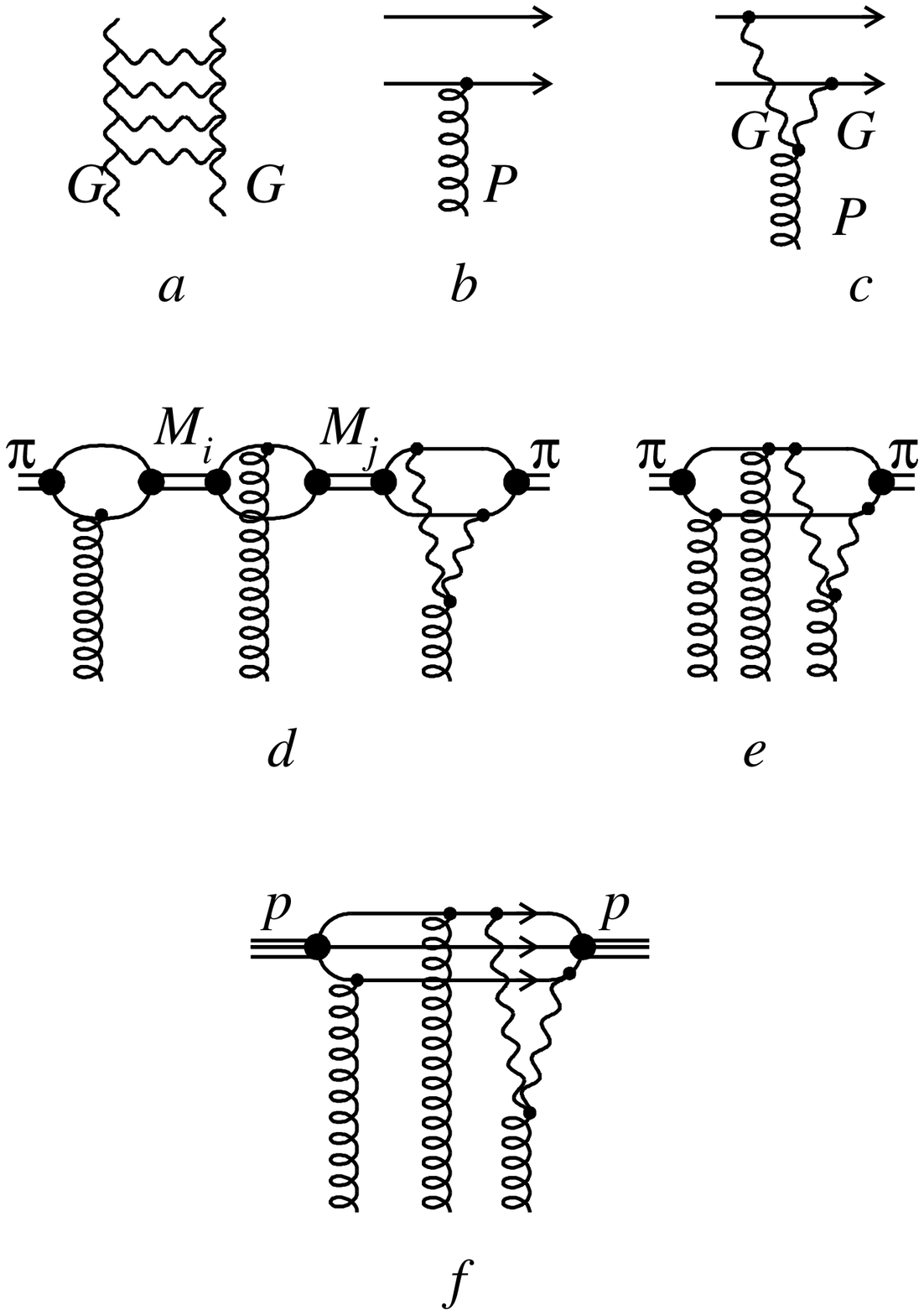}}
\end{center}
Fig. 1. $a)$ Gluon ladder diagram of the Lipatov's pomeron;
 $b,c)$ various types of  pomeron-quark couplings; $d)$ diagrams
of soft multiple rescatterings for the pion beam
written using the language of hadron
intermediate states; the summation is performed
over all allowed meson states
$M_i$ and $M_j$;
$e)$ the same diagrams as in fig. 1$d$ but written
in the language of quark rescatterings;
$f)$ the same as Fig. 1$e$, but for the proton.
\end{figure}

\begin{figure}
\begin{center}
\mbox{\epsfxsize=13cm \epsfbox{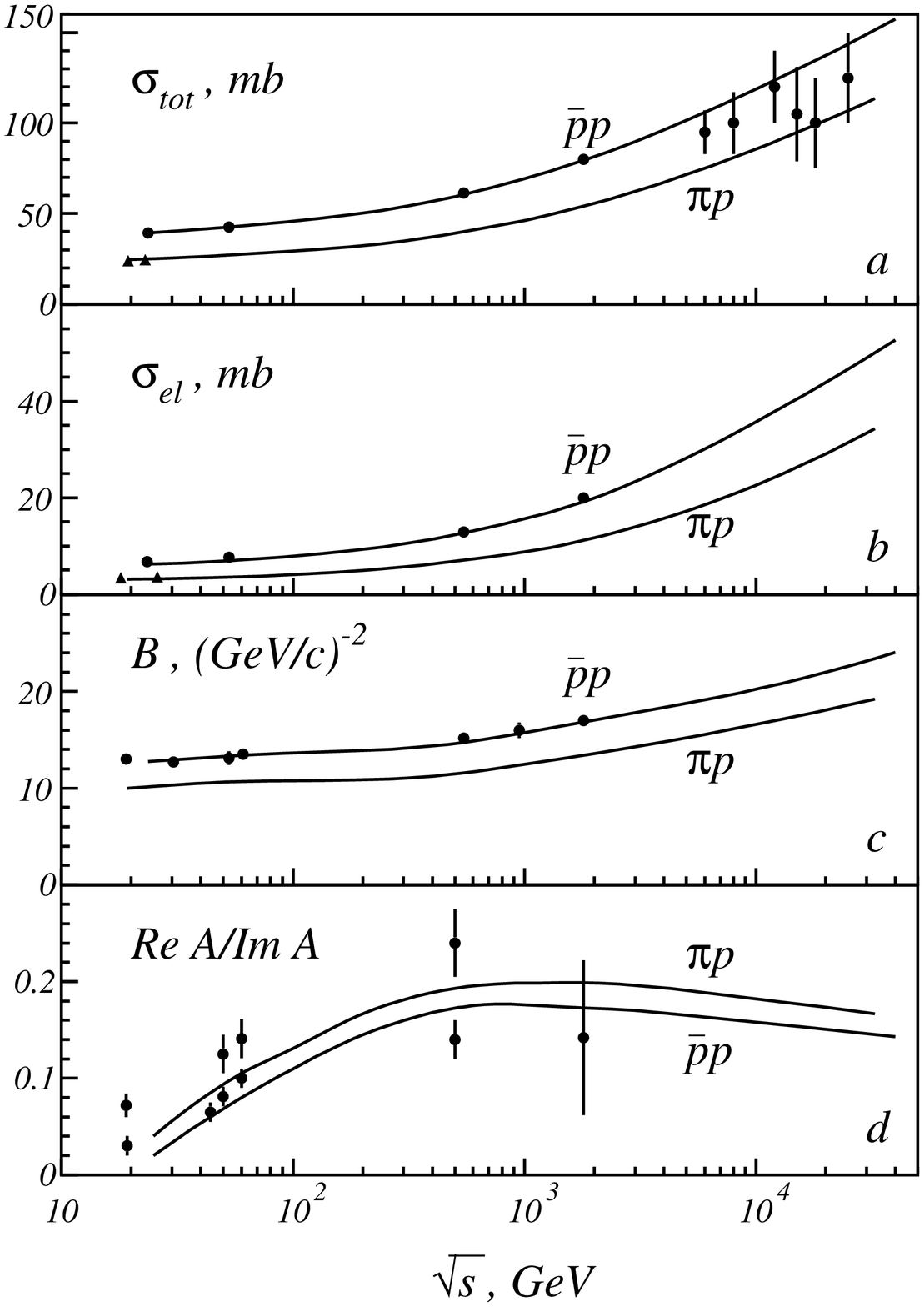}}
\end{center}
Fig. 2.  Experimental data and calculated values for $pp(\bar pp)$
and $\pi p$ collisions: $a$) total cross sections, the data at $\sqrt {s}>
5000$ GeV are Akeno cosmic ray experiment; $b$) elastic cross sections;
$c$) slope parameter the elastic scattering; $d$) $\rho=
Re\,A/Im\,A$ for $pp(\bar pp)$ and $\pi p$ scattering.
\end{figure}

\begin{figure}
\begin{center}
\mbox{\epsfxsize=10cm \epsfbox{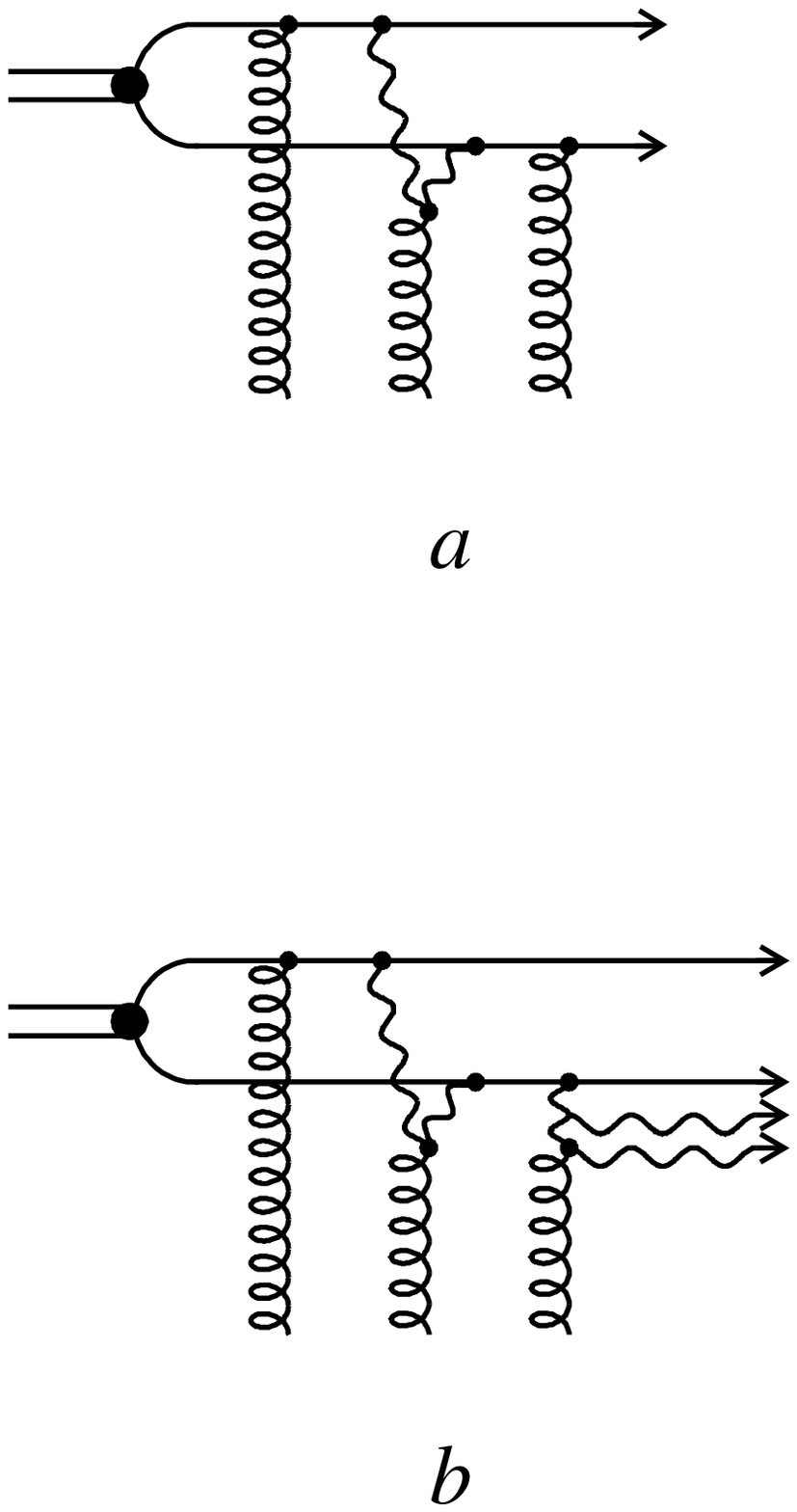}}
\end{center}
Fig. 3. Examples of DD processes which are measured in
the experiments: $a)$ The dissociation of hadron, $b)$ Process with
the partly dissociated pomeron - this process has not been taken into
account in eqs. (3)-(4).
\end{figure}

\begin{figure}
\begin{center}
\mbox{\epsfxsize=13cm \epsfbox{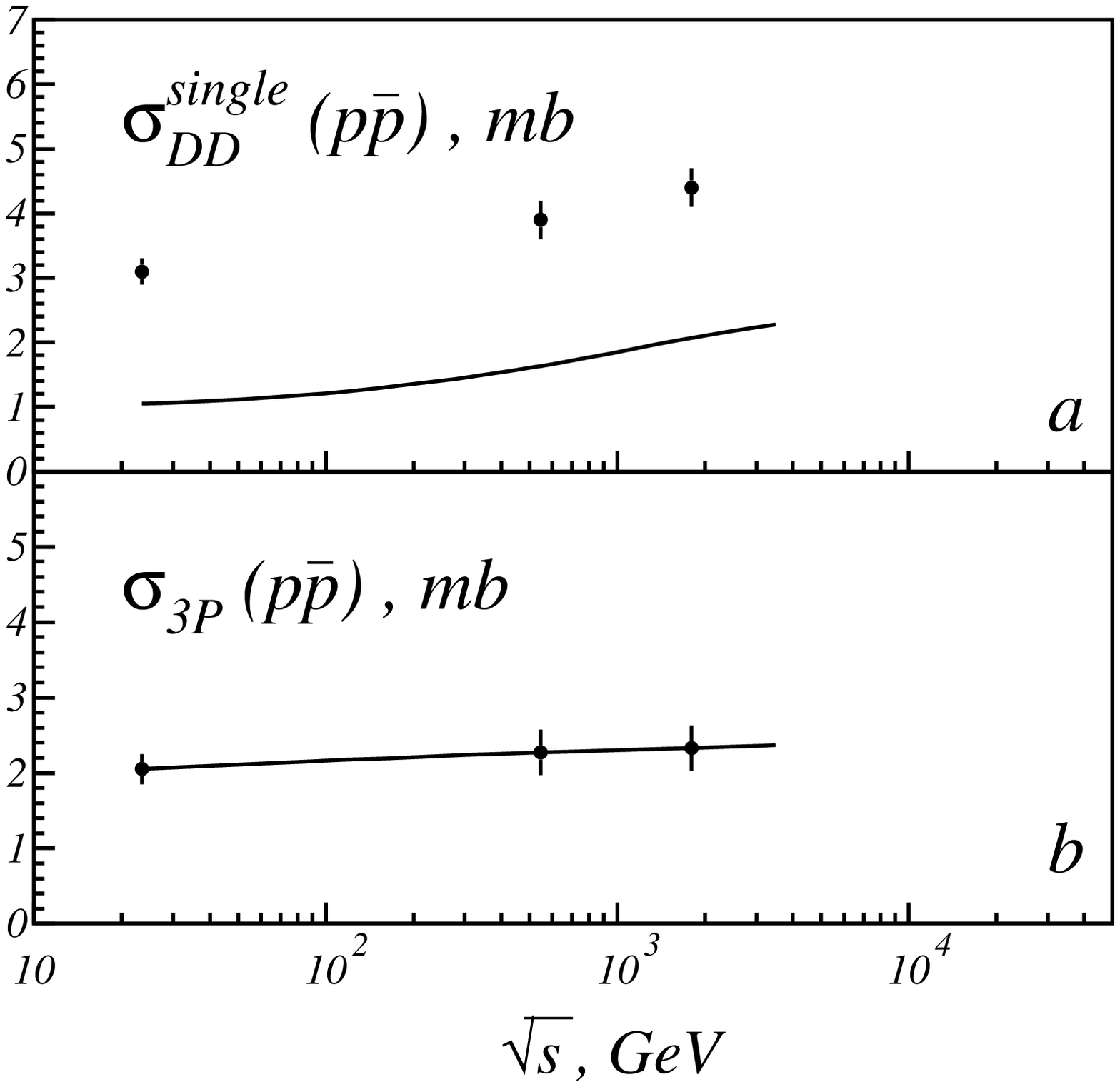}}
\end{center}
Fig. 4. $a)$ Experimental
data for the proton diffraction dissociation
in $p\bar p$ collisions [13] measured as $\sigma_{DD}(exp)=\int dM^2
d\sigma/dtdM^2,\;0<-t<0.4,\; M^2/s<0.15$; curve presents the  values  of
$\sigma^{single}_{DD}(pp)$ calculated by eq. (3).
$b$) Cross section of the diffraction  dissociation of the pomeron
evaluated as $\sigma_{3P}(pp)=\sigma_{DD}(exp)-\sigma^{single}_{DD}(pp)$;
curve is a parametrization  of the three-pomeron diffractive
cross section in the form $(1.85+0.033\ln s/s_0)$ mb.
\end{figure}

\begin{figure}
\begin{center}
\mbox{\epsfxsize=13cm \epsfbox{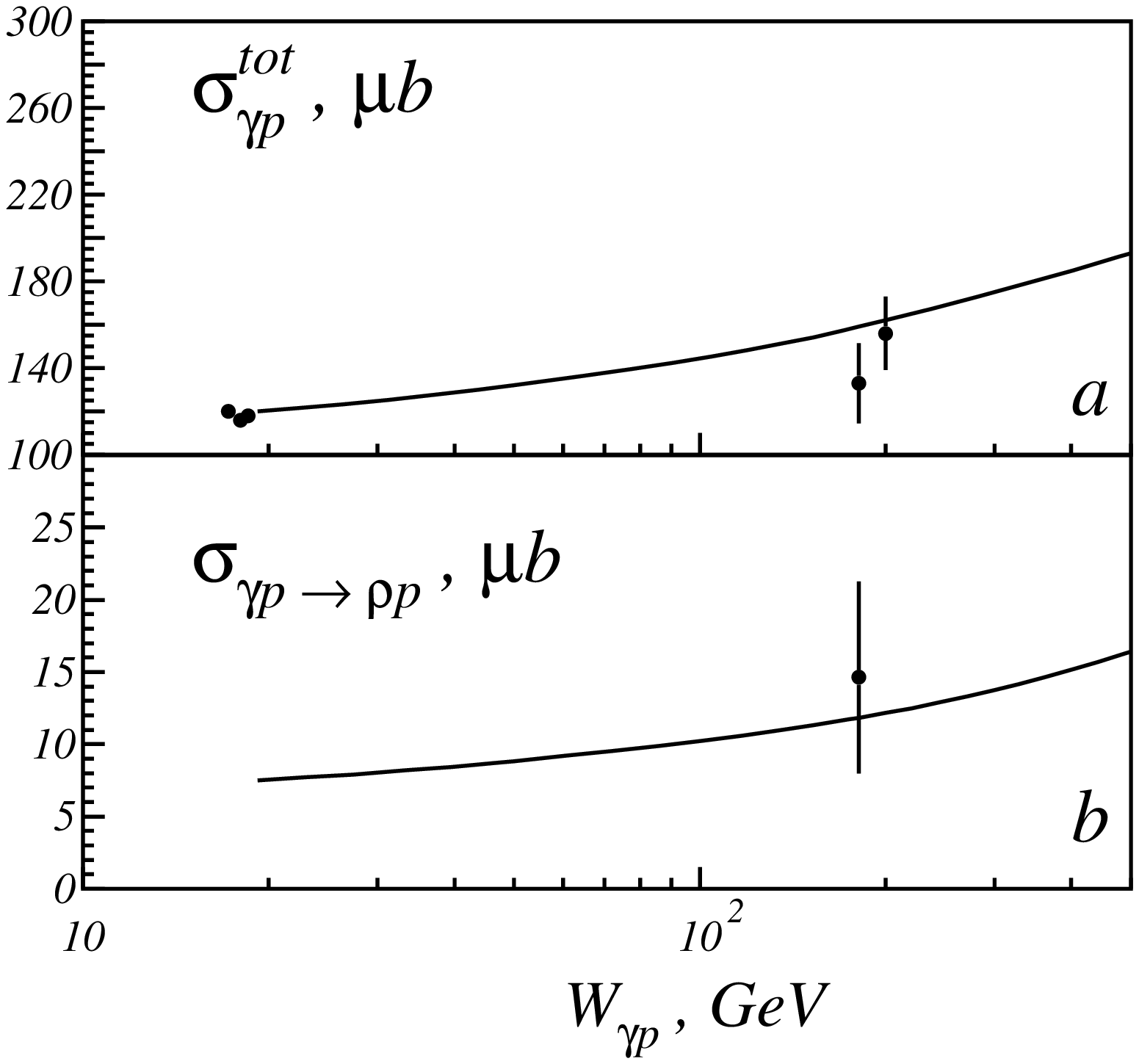}}
\end{center}
Fig. 5. a) Total cross section $\sigma_{tot}(\gamma p)$ calculated
by eq. (1) (solid curve) and b) cross section for the two-particle
reaction $\sigma_{tot}(\gamma p \to \rho p)$
calculated by eq. (2).
\end{figure}

\begin{figure}
\begin{center}
\mbox{\epsfxsize=13cm \epsfbox{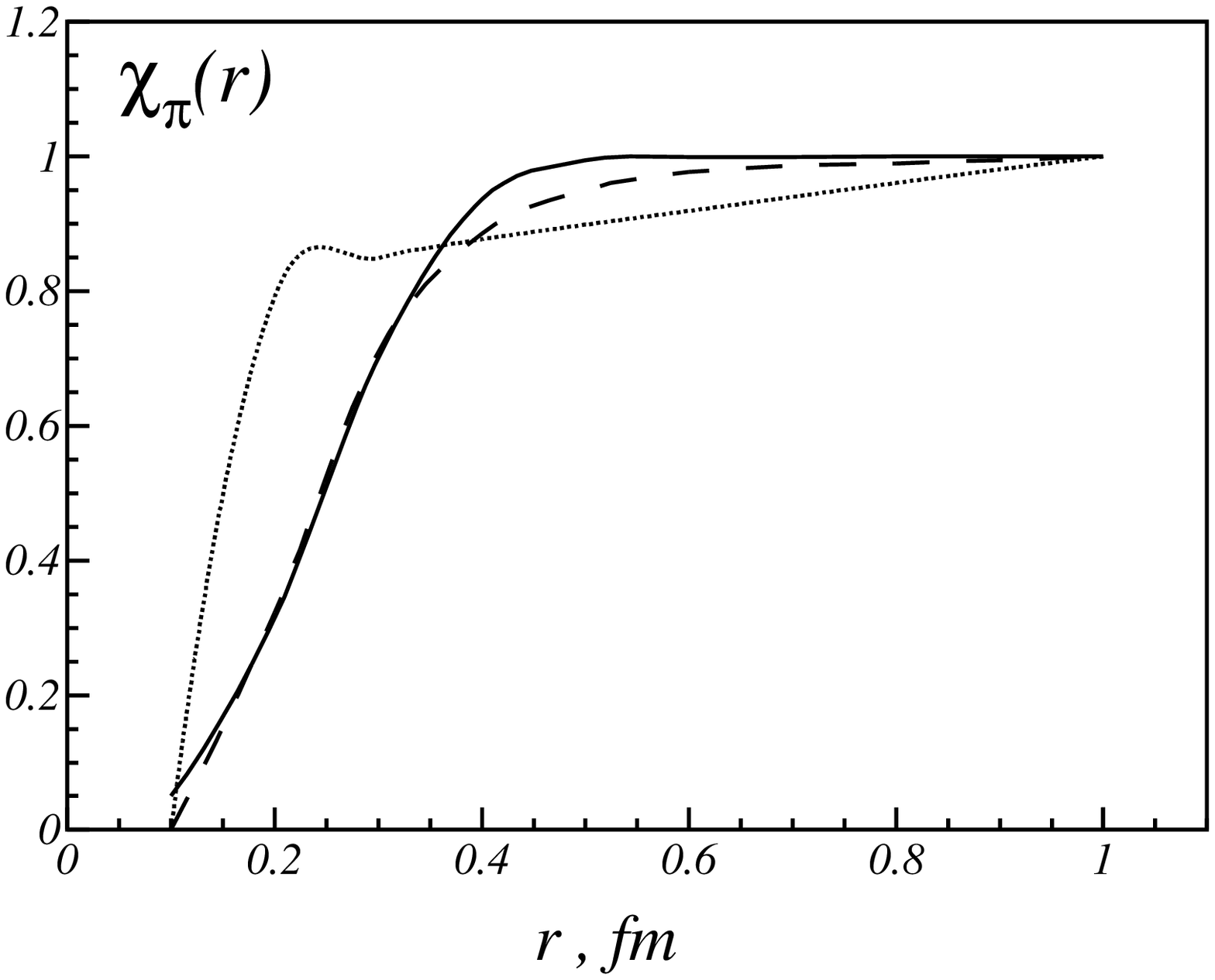}}
\end{center}
Fig. 6. Colour screening profile factor $\chi_{\pi}(r)$ as a function
of the interquark distance (for pion-proton collision).
Dashed curve stands for the energy  $\sqrt{s}=546$ GeV;
 dotted curve is calculated at far
asymptotic energies $s=10^{30}$ GeV$^2$; solid curve stands for
 $\chi_{\pi}(r)$ parametrized in the form
$1-\exp(-r^n/r_{cs,eff}^n)$ with $n=2.86$ and $r_{cs,eff}=0.281$ fm.
\end{figure}

\begin{figure}
\begin{center}
\mbox{\epsfxsize=13cm \epsfbox{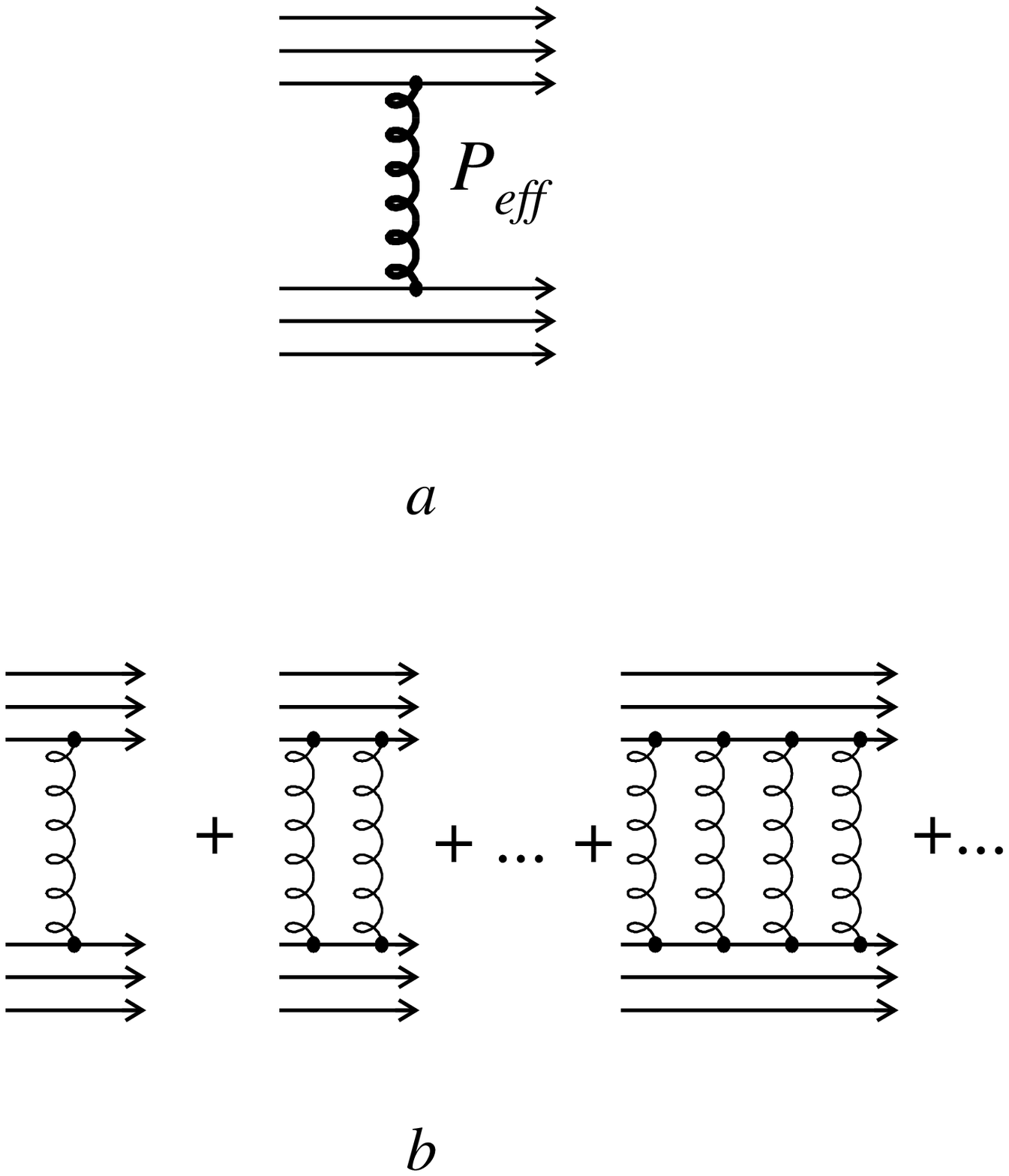}}
\end{center}
Fig. 7. Impulse approximation diagrams:
$a$) Exchange of the effective pomeron in the $pp$-collision which was
used in ref. [9]; $b$) Eikonal approximation diagrams for the quark-quark
interaction which play the role of the effective pomeron at moderately
high energies.
\end{figure}

\begin{figure}
\begin{center}
\mbox{\epsfxsize=13cm \epsfbox{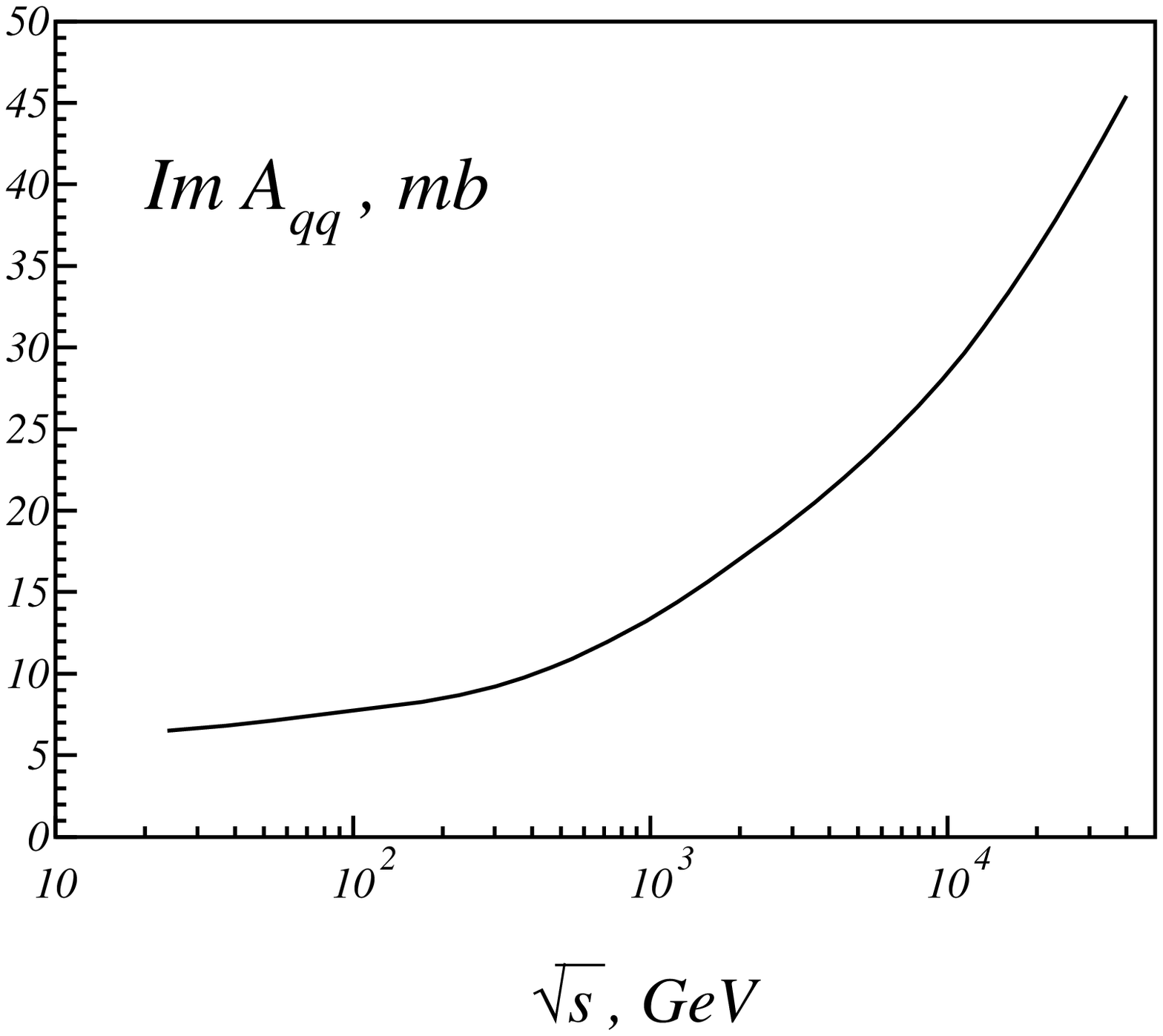}}
\end{center}
Fig. 8. Energy dependence of the cross section related to the
set of diagrams shown in  fig. 7$b$.
\end{figure}


\begin{thebibliography} {aa} \fussy
\bibitem{1} L.N.Lipatov, Sov.Phys. JETP  {\bf 63}, 904, 1986.
\bibitem{2}  A.Capella, J.Tranh Thanh Van, and J.Kwiecinski,
Phys. Rev. Lett. {\bf 58}, 2015, 1987;\\
  M.G.Ryskin and Yu.M.Shabelski, Z. Phys. {\bf C56}, 253, 1992;\\
 P.V.Landshoff,
"Soft Hadron Reactions", in "QCD, 20 Years Later", vol.1,\\
ed. P.M.Zervas and H.A.Kastrup, World Scientific, Singapore, 1993;\\
 R.S.Fletcher, T.K.Gaisser, and F.Halzen, Phys. Lett., {\bf B298}, 442,
 1993;\\
 E.Gotsman, E.M.Levin, and U.Maor, Phys. Rev. {\bf D49}, R4321, 1994.
\bibitem{3} E.A.Kuraev, L.N.Lipatov, and V.S.Fadin, Sov.Phys. JETP  {\bf 44},
 443, 1976;\\
 Ya.Ya.Balitsky and L.N.Lipatov, Sov.J.Nucl.Phys. {\bf 28}, 882, 1978.
\bibitem{4}  A.D.Martin, W.J.Stirling, and R.G.Roberts,
Phys. Rev. {\bf D50}, 6734, 1994.
\bibitem{5}  ZEUS Collaboration: M.Rocco, Proc. of 29th
Rencontre de Moriond,
March 1994, ed. J.Tran Thanh Van;\\
 H1-Collaboration: K.Mueller, Proc. of 29th Rencontre de Moriond,
March 1994, ed. J.Tran Thanh Van.
\bibitem{6}  A.J.Askew, J.Kwiecinski, A.D.Martin, and P.J.Sutton,
Phys. Rev. {\bf D47}, 3775, 1993; ibid, {\bf D49}, 4402, 1994.
\bibitem{7}  V.V.Anisovich, M.N.Kobrinsky, J.Nyiri and Yu.M.Shabelski,
"Quark Model and High Energy Collisions", World Scientific,
Singapore, 1985, \\
 P.V.Landshoff and O.Nachtman, Z. Phys.
{\bf C35}, 405, 1987.
\bibitem{8}  T.K.Gaisser and T.Stanev, Phys. Lett., {\bf B219}, 375, 1989;\\
 M.Block, F.Halzen and B.Margolis, Phys. Lett., {\bf B252}, 481, 1990;\\
R.S.Fletcher, Phys. Rev. {\bf D46}, 187, 1992.
\bibitem{9}  V.V.Anisovich, L.G.Dakhno and V.A.Nikonov,
Phys. Rev. {\bf D44}, 1385, 1991.
\bibitem{10}  A.Donnachie and P.V.Landshoff, Nucl. Phys. {\bf B231},
189, 1984;\\
A.B.Kaidalov and K.A.Ter-Martirosyan, Sov. J. Nucl. Phys.
{\bf 39}, 979, 1984.
\bibitem{11}  V.V.Anisovich and L.G.Dakhno, Nucl.Phys. (Proc. Suppl.)
{\bf 25B}, 247, 1992.
\bibitem{12}  L.N.Lipatov, Phys. Lett. {\bf B251}, 284, 1990;
ibid, {\bf B303}, 394, 1993;\\
  J.Bartels, Phys. Lett. {\bf B298}, 204, 1993 and
 "Unitarity Correction to the Lipatov
Pomeron", Preprint DESY 93-028, 1993;\\
J.Bartels and M.G.Ryskin, Z. Phys. {\bf C56}, 1751, 1993.
\bibitem{13}  M.G.Albrow et al.,Nucl. Phys. {\bf B108}, 1, 1976;\\
S.Belforte, G.Chiarelli, P.Giromini, S.Miscetti and
R.Paoletti, "Measurement of Small Angle Antiproton-Proton Elastic
Scattering at $\sqrt {s}=546$ and $\sqrt {s}=1800$ GeV",
Preprint CDF/ANAL/CDF/CDFR 2049, 1994;\\
 S.Belforte, G.Chiarelli, P.Giromini, K.Goulianos, S.Miscetti and
R.Paoletti, "Measurement of $\bar p p$ Single Diffraction Dissociation
 at $\sqrt {s}=546$ and $\sqrt {s}=1800$ GeV",
Preprint CDF/ANAL/CDF/CDFR 2050, 1994;\\
S.Belforte, G.Chiarelli, P.Giromini, S.Miscetti and
R.Paoletti, " Measurement of the Anti\-proton-Proton Total Cross
Section at $\sqrt {s}=546$ and $\sqrt {s}=1800$ GeV",
Preprint\\ CDF/ANAL/CDF/CDFR 2051, 1994.
\bibitem{14}  J.Pumplin, Phys.Rev. {\bf D8}, 2899, 1973.
\bibitem{15}  N.A.Pavel (ZEUS collaboration), "New results from ZEUS",
Preprint DESY 93-160, 1993;\\
G.Wolf, "HERA Physics", Preprint DESY 94-022, 1994.

\bibitem{16} V.V.Anisovich, L.G.Dakhno, V.A.Nikonov and M.G.Ryskin,
    Phys. Lett. {B292}, 169, 1992;\\
  V.V.Anisovich, L.G.Dakhno and M.M.Giannini,
    Phys. Rev. {\bf C49}, 3275, 1994.

\bibitem{17}  G.P.Lepage, J.Comp. Phys. {\bf 27}, 192, 1978.
\end{thebibliography}
\end{document}